\documentclass[aps,prb,twocolumn,showpacs,superscriptaddress,amsmath,amssymb,citeautoscript]{revtex4-1}

\usepackage{graphicx}
\usepackage{bm}
\renewcommand{\vec}[1]{\bm{#1}}

\begin{document}


\title{Double zigzag spin chain in strong magnetic field close to saturation }



\author{I. T. Shyiko}
\affiliation{Institute of High Technologies, Taras Shevchenko
  National University of Kiev, 03022 Kiev, Ukraine}

\author{I. P. McCulloch} 
\affiliation{School of Physical Sciences, The
  University of Queensland, Brisbane, QLD 4072, Australia }

\author{J. V. Gumenjuk-Sichevska}
\affiliation{V.~E.~Lashkarev Institute of Semiconductor Physics, National Academy of Sciences, 03028 Kiev, Ukraine}

\author{A. K. Kolezhuk}
\affiliation{Institute of High Technologies, Taras Shevchenko
  National University of Kiev,  03022 Kiev, Ukraine}
\affiliation{Institute of Magnetism, National Academy of Sciences and Ministry of Education,  03142 Kiev, Ukraine}

\date{\today}

\begin{abstract}
We study the ground state phase diagram of a frustrated spin tube in a
strong external magnetic field.  This model can be viewed as two coupled zigzag
spin chains, or as a two-leg spin ladder with frustrating
next-nearest-neighbor couplings along the legs, and its study is
motivated by the physics of such materials as Sulfolane-$\rm
Cu_{2}Cl_{4}$ and $\rm BiCu_{2}PO_{6}$.  In magnetic fields right below the
saturation, the system can be effectively represented as a
dilute gas of two species of bosonic quasiparticles that correspond to
magnons with inequivalent incommensurate momenta at two degenerate
minima of the magnon dispersion. Using the method previously proposed
and tested for frustrated spin chains, we calculate effective
interactions in this two-component Bose gas. On this basis, we establish the phase
diagram of nearly-saturated frustrated spin tube, which is shown to 
include the two-component Luttinger liquid, two types of vector
chiral phases, and phases whose physics is determined by the presence
of bound magnons.  We study the phase diagram of the model numerically by means of the
density matrix renormalization group technique, and find 
a good agreement with our analytical predictions.
\end{abstract}



\maketitle

\section{Introduction}
\label{sec:intro}

Frustrated spin systems, especially in low dimensions, display a
rich variety of unconventionally ordered ground states 
\cite{Diep2004book,Mila2011book}. Strong external magnetic field,
competing with the exchange interaction, can serve as 
a control parameter that drives the corresponding quantum phase transitions.
The ground state of a
frustrated quantum spin system is 
considerably simplified in a sufficiently
strong external field that eventually leads to a
fully polarized state above some critical field value $H_{s}$
(strictly speaking, the latter
is true only for an axially-symmetric case, but we assume that
deviations from axial symmetry are negligibly small). In fields just slightly
below $H_{s}$,  one may view the system as a dilute gas of
excitations (magnons) on
top of the fully polarized state
\cite{BatyevBraginskii84,Johnson86,Gluzman94,NikuniShiba95,Okunishi98,JackeliZhitomirsky04,UedaTotsuka09}. At
low density of magnons they can be approximately treated as bosonic quasiparticles.
In the case of a strong frustration, the magnon dispersion has two or more degenerate minima at
inequivalent incommensurate wave vectors, so one arrives at the picture
of a multicomponent dilute Bose gas.
In the one-dimensional case, infrared singularities, appearing in the description of effective
interactions in the magnon gas, require special treatment  \cite{Kolezhuk+12}.

 Depending on the ratio of interactions
between the same or different sorts of particles, several types of the
ground state can be favored.
Particularly, in two- and three-dimensional systems, different kinds
of helical order (``fan'' and ``umbrella'') are realized
\cite{NikuniShiba95,UedaTotsuka09}, while in
one dimension  quantum
fluctuations destroy long-range helical order and may lead to the
formation of several different states with competing types of
unconventional short- and long-range orders.
In one dimension, in the case of repulsion between magnons, the
 ``umbrella''  and ``fan'' phases get replaced by the vector chiral
(VC) long-range order
\cite{KV05,Mc+08,Okunishi08,KMc09,Hikihara+10} (which is equivalent to the local
spin current) and by 
the two-component Tomonaga-Luttinger liquid (TLL2)\cite{Okunishi+99},
respectively. 
On the other hand, attraction between quasiparticles can lead
to the appearance of a short-range multipolar (spin nematic) order
\cite{Hikihara+08,Sudan+09}, or alternatively to metamagnetic jumps \cite{Arlego+11}.

\begin{figure}[tb]
 \includegraphics[width=0.33\textwidth]{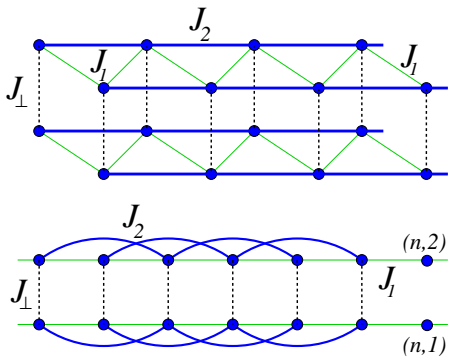}
 \caption{
\label{fig:tube} (Color online).
The frustrated spin tube described by the Hamiltonian
(\ref{hamS}). The tube can be alternatively viewed as two zigzag
chains coupled by the transversal interaction $J_{\perp}$, or as a spin ladder with
 next-nearest-neighbor exchange couplings along the legs (lower panel).}
\end{figure}

Recently, the above approach, based on the mapping to the multicomponent Bose gas, has been successfully
applied to spin-$S$ zigzag spin chain
\cite{Arlego+11,Kolezhuk+12}, which is a paradigmatic model of a
frustrated spin system. It has been shown that for zigzag chains close to
saturation  this approach is
able to capture the physics of phase transitions between the VC and
TLL2 phases, and for $S\geq 1$ it can detect the boundary of the
(metamagnetic) region where bound states of magnons are formed.  In
the present paper, we employ this method to study the
strong-field part of the ground state phase diagram of the frustrated
spin tube shown in Fig.\ \ref{fig:tube}.
The spin tube, which will be the subject of our study, can
be viewed as two coupled zigzag spin chains, or as a two-leg spin
ladder with frustrating next-nearest-neighbor couplings along the
legs, see Fig.\ \ref{fig:tube}.  This model is described by the
following Hamiltonian:
\begin{eqnarray} 
\label{hamS}
\mathcal{H}&=& \sum_{n=1}^{L}\sum_{a=1,2} \Big\{ J_{\perp}
(\vec{S}_{n,1}\cdot\vec{S}_{n,2}) -H(S_{n,1}^{z}+S_{n,2}^{z})\\
& +& J_{1} (\vec{S}_{n,a}\cdot\vec{S}_{n+1,a}) +
J_{2} (\vec{S}_{n,a}\cdot\vec{S}_{n+2,a})  \Big\},\nonumber 
\end{eqnarray}
where $\vec{S}_{n,a}$ are spin-$S$ operators acting at the $n$-th site
of the $a$-th leg,
 $J_{1}$  and $J_{2}$  are the nearest-neighbor (NN) and
next-nearest neighbor (NNN) exchange couplings along the legs, $J$ is
the rung exchange,
and $H$ is the external magnetic field. The system may be alternatively viewed
as four antiferromagnetic chains connected by  rung and zigzag couplings.
This model has
been recently studied at zero field
\cite{VekuaHonecker06,Lavarelo+11}. It is believed to be relevant for the physics of
such quasi-one-dimensional materials as Sulfolane-$Cu_{2} Cl_{4}$ ($\rm Cu_{2} Cl_{4} H_{8} C_{4}
SO_{2}$) 
\cite{Fujisawa+03}, which exhibited unusual critical behavior in the
field-induced transition from a helimagnetic to a non-magnetic phase \cite{Garlea+08,Garlea+09,Zheludev+09},
and  $\rm
BiCu_{2}PO_{6}$, which has attracted the attention of several research
groups as being a realization of a frustrated ladder system with
incommensurate correlations
\cite{Koteswararao+07,Mentre+09,Tsirlin+10}.  Apart from the possible
relevance for the above materials, this model is fundamentally interesting since it presents the simplest
example of two interacting zigzag chains.

In this paper, we are
interested in the frustrated case, so $J_{2}$ is chosen to be positive
 while $J_{1}$ may have any sign. 
It is convenient to use the quantity
\begin{equation} 
\label{beta} 
\beta=J_{1}/J_{2}
\end{equation}
as the frustration parameter. We will be interested in the regime
$|\beta|<4$,  when
the magnon dispersion develops two degenerate minima
at inequivalent points $\pm Q$ in the momentum space  
(in what follows, we refer to this regime as ``strong frustration'').

For $S=\frac{1}{2}$, the phase diagram of the above model in the
absence of the magnetic field has been studied numerically
\cite{Lavarelo+11} and was found to contain the rung singlet and the
columnar dimer phase.  Earlier, a slightly different version of the
model including exchange coupling along diagonals of the ladder has
been investigated \cite{VekuaHonecker06}; its phase diagram at zero
field has been shown to contain the rung singlet phase, the Haldane
phase, and two different columnar dimerized phases. The magnetic phase
diagram of both versions of the model is at present unexplored.

We study the ground state of the strongly frustrated ($|\beta|<4$)
spin-$S$ tube decribed by the Hamiltonian (\ref{hamS}) in high magnetic fields in the immediate
vicinity of saturation, for spin values $S=1$ and $S=\frac{1}{2}$.  It
is shown that the phase diagram contains the two-component Luttinger
liquid, two types of vector chiral phases, and phases whose physics is
determined by the presence of bound magnons.  We compare our
analytical predictions with the results of numerical simulations using
the density matrix renormalization group\cite{white92,schollwoeck05}
(DMRG) technique.  To that end, we compute the chirality correlation
function and magnetization distribution as functions of $\beta$ at
several values of $J_{\perp}$, at fixed magnetization close to
saturation. We demonstrate that the DMRG results are in a good agreement with our theoretical predictions.

The structure of the paper is as follows: in
Sect.\ \ref{sec:efftheory} we  describe the mapping of the spin tube problem to the dilute
two-component lattice Bose gas and outline the main steps of computing
effective interactions. 
Section \ref{sec:tube-results} discusses the specific predictions of the theory
for spin tubes  with $S=\frac{1}{2}$ and $1$,
while Sect.\ \ref{sec:numerics} presents the
results of numerical analysis and their comparison with analytical
predictions. Finally, Sect.\ \ref{sec:summary} contains a brief
summary.

\section{Effective  two-component Bose gas description of the spin tube}
\label{sec:efftheory}

We intend to  map the spin problem (\ref{hamS}) to a dilute gas of
interacting magnons,  for values of the field $H$ slightly lower than the
saturation field $H_{s}$. For that purpose,
it is convenient to use the Dyson-Maleev representation  for the
spin operators in  (\ref{hamS}):
\begin{eqnarray} 
\label{DysonMaleev} 
&& S_{nm}^{+}=\sqrt{2S}b_{nm},\quad S_{nm}^{-}=\sqrt{2S}b_{nm}^{\dag} 
\Big(1-\frac{b_{nm}^{\dag}b_{nm}^{\vphantom{\dag}}}{2S}\Big),\nonumber\\
&& S_{nm}^{z}=S-b_{nm}^{\dag}b_{nm}^{\vphantom{\dag}},
\end{eqnarray}
where $b_{nm}$ are bosonic operators acting at site $(nm)$ of the
lattice, and $n=1,\ldots,L$
 and $m=1,2$
denote the rung and leg numbers, respectively, see
Fig.\ \ref{fig:tube}. 
To enforce the constraint
$b_{nm}^{\dag}b_{nm}^{\vphantom{\dag}}\leq 2S$, one can add 
the infinite interaction term
to the
Hamiltonian, which reads:
\begin{equation} \label{infU} 
\mathcal{H}\mapsto
\mathcal{H}+U\sum_{nm}:(b_{nm}^{\dag}b_{nm}^{\vphantom{\dag}})^{2S+1}:\;,\quad U\to+\infty,
\end{equation}
where $:(\ldots):$ denotes normal ordering. At the
level of two-body interactions (which are dominating because of the
diluteness of the gas) this term should be taken into account only for $S=\frac{1}{2}$.

Assuming periodic boundary conditions, we pass to the momentum representation for bosonic operators, 
\[
b_{nm}=\frac{1}{\sqrt{2L}} \sum_{\vec{k}}b_{\vec{k}}e^{ik_{x}n+k_{y}m},
\]
where $\vec{k}=(k_{x},k_{y})$, with $k_{y}$ taking only values $0$ or
$\pi$, and $L$ is the total number of rungs.
Then one can cast the
Hamiltonian (\ref{hamS}) in the following form:
\begin{equation}
\label{hamB} 
\mathcal{H}=\sum_{\vec{k}} E_{\vec{k}} b^{\dag}_{\vec{k}}b_{\vec{k}}^{\vphantom{\dag}}
+\frac{1}{4L}\sum_{\vec{k}\vec{k'}\vec{q}}V_{\vec{q}}(\vec{k},\vec{k'})b^{\dag}_{\vec{k}+\vec{q}}b^{\dag}_{\vec{k'}-\vec{q}}
b_{\vec{k}}^{\vphantom{\dag}} b_{\vec{k'}}^{\vphantom{\dag}}.
\end{equation}
Here the magnon dispersion
$E_{\vec{k}}$ is given by
\begin{equation} 
\label{ek} 
E_{\vec{k}}=H-2(J_{1}+J_{2})S+2SJ_{k_{x}} +J_{\perp}S(\cos k_{y}-1),
\end{equation}
where we use the shorthand notation
\begin{equation} 
\label{shorthand}
J_{k}\equiv J_{1}\cos(k)+ J_{2}\cos(2k). 
\end{equation}
In the case of the strong frustration $|\beta|<4$, which is
of main interest for us, there are two  inequivalent degenerate minima
of $E_{\vec{k}}$ that are reached at wave vectors $\vec{k}=(\pm Q,\pi)$ and
$\vec{k}=(\pm Q,0)$ for positive and negative $J_{\perp}$,
respectively. The wave vector $Q$ is incommensurate and is given by 
\begin{equation} 
\label{Q} 
Q=\arccos(-\beta/4).
\end{equation}
The saturation field $H_{s}$ can be found \cite{noteHs}  from the condition 
$(\min \left. E_{\vec{k}})\right|_{H=H_{s}}=0$ :
\begin{equation} 
\label{Hs} 
H_{s}=2S\big\{J_{1}+J_{2} +J_{\perp}\theta(J_{\perp}) -J_{Q}\big\},
\end{equation}
where $\theta(x)$ is the Heaviside function.
The external field may be viewed as playing the role of the chemical potential
$\mu=H_{s}-H$ for magnons. In what follows, it is convenient to introduce instead
of $E_{\vec{k}}$ the
quantity  
\begin{equation} 
\label{eps-k} 
\varepsilon_{\vec{k}}=E_{\vec{k}}+\mu.
\end{equation}

For  $S\geq 1$, the two-body interaction $V_{\vec{q}}(\vec{k},\vec{k'})$ depends on
the transferred momentum $q$ as well as on the incoming momenta $k$, $k'$:
\begin{eqnarray} 
\label{Vq} 
V_{\vec{q}}(\vec{k},\vec{k'})&=&2J_{q_{x}}-J_{k_{x}}-J_{k_{x}'}\\
&+&J_{\perp}\big\{
\cos q_{y} -\frac{1}{2}(\cos k_{y} +\cos k_{y}') \big\}.\nonumber
\end{eqnarray}
For spin $\frac{1}{2}$, one has to add the term (\ref{infU}) to the
Hamiltonian, simultaneously dropping the terms  like
$b_{nm}b_{nm}$ involving double occupancy. As a result, for
$S=\frac{1}{2}$ the expression
for the two-body interaction simplifies to
\begin{equation} 
\label{Vq-s12} 
V_{\vec{q}}(\vec{k},\vec{k'})=U+J_{\perp}\cos q_{y} +2J_{q_{x}}, \quad
U\to +\infty.
\end{equation}

\begin{figure}[tb]
 \includegraphics[width=0.4\textwidth]{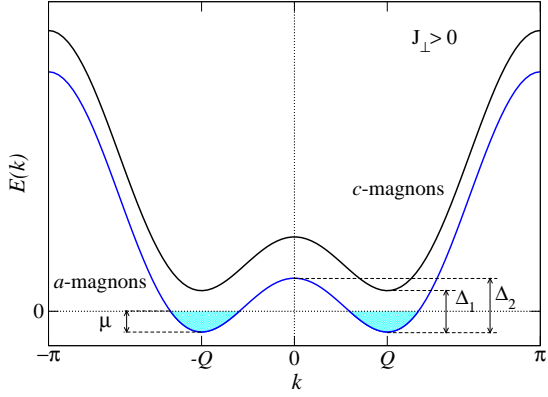}
 \caption{
\label{fig:ac-magnons} (Color online).
Schematic picture of the magnon dispersion in the vicinity of the
saturation field. For the sake of definiteness, the case of
antiferromagnetic $J_{\perp}$ is shown. The system is populated by $a$-magnons
with the momenta close to $\pm Q$, provided that the chemical
potential $\mu=H_{s}-H$ satisfies the conditions (\ref{vicinity}) and (\ref{vicinity-2}).
}
\end{figure}

The
model (\ref{hamB}) describes two magnon branches of different parity
with respect to the permutation of the ladder legs. We denote
the operators describing even and odd magnons by $c_{k}$ and $a_{k}$,
respectively:
\begin{equation} 
\label{ac} 
a_{k}=b_{(k,\pi)},\quad c_{k}=b_{(k,0)}.
\end{equation}
In this notation, the Hamiltonian takes the form
\begin{widetext}
\begin{eqnarray}
\label{hamB-ac} 
\mathcal{H}&=&\sum_{\vec{k}} \Big\{ (\varepsilon_{k}^{a}-\mu)
a^{\dag}_{k}a_{k}^{\vphantom{\dag}} +(\varepsilon_{k}^{c}-\mu)
c^{\dag}_{k}c_{k}^{\vphantom{\dag}}\Big\}
+\frac{1}{2L}\sum_{kk'q}\Big\{ 
V_{q}^{aa}(k,k') a^{\dag}_{k+q} a^{\dag}_{k'-q} a_{k}^{\vphantom{\dag}} a_{k'}^{\vphantom{\dag}}
+V_{q}^{cc}(k,k') c^{\dag}_{k+q} c^{\dag}_{k'-q}
c_{k}^{\vphantom{\dag}} c_{k'}^{\vphantom{\dag}} \nonumber\\
&+&V_{q}^{ac}(k,k') a^{\dag}_{k+q} a^{\dag}_{k'-q} c_{k}^{\vphantom{\dag}} c_{k'}^{\vphantom{\dag}}
+V_{q}^{ca}(k,k') c^{\dag}_{k+q} c^{\dag}_{k'-q} a_{k}^{\vphantom{\dag}} a_{k'}^{\vphantom{\dag}}
+V_{q}^{\times}(k,k') c^{\dag}_{k+q} a^{\dag}_{k'-q}
c_{k}^{\vphantom{\dag}} a_{k'}^{\vphantom{\dag}} \nonumber
\Big\},
\end{eqnarray}
\end{widetext}
The magnon energies and interaction amplitudes above can be read off
 Eqs.\ (\ref{ek}), (\ref{eps-k}), (\ref{Vq}), (\ref{Vq-s12}).  For
 the energies, one has 
\begin{eqnarray} 
\label{e-ac} 
&& \varepsilon_{q}^{a}=2S(J_{q}-J_{Q}),\;\;
\varepsilon_{q}^{c}=\varepsilon_{q}^{a}+2J_{\perp}S \quad \text{at
  $J_{\perp}>0$},\\
&& \varepsilon_{q}^{c}=2S(J_{q}-J_{Q}),\;\;
\varepsilon_{q}^{a}=\varepsilon_{q}^{c}+2|J_{\perp}|S \quad \text{at  $J_{\perp}<0$},\nonumber
\end{eqnarray}
thus the energies of $a$-branch lie below (above) those of the
 $c$-branch for $J_{\perp}>0$ ($J_{\perp}<0$),
respectively. 
When the magnetic field is decreased below
$H_{s}$, the ground state of the system can be viewed as a dilute gas
of $a$-magnons for $J_{\perp}>0$ or of $c$-magnons for
$J_{\perp}<0$. Therefore, the last term in (\ref{hamB-ac}), describing
the scattering of a $c$-magnon on an $a$-magnon, does
not influence the structure of the ground state \emph{in the immediate
vicinity of the saturation field}: under the condition
\begin{equation} 
\label{vicinity}
\mu=H_{s}-H \ll \Delta_{1}\equiv 2SJ_{\perp}
\end{equation} 
(see Fig.\ \ref{fig:ac-magnons}) there is simply no regime when densities of both sorts of magnons ($a$
and $c$) are \emph{simultaneously} nonzero. In what follows, we 
assume that the condition (\ref{vicinity}) is always satisfied, so we
can safely ignore the presence of the last term in
(\ref{hamB-ac}). However, the other amplitudes in (\ref{hamB-ac}),
e.g., describing conversion of a pair of $a$-magnons to a pair of
$c$-magnons, have to be kept, because they contribute to intermediate virtual
states in multiple scattering processes.

The other two-body interaction amplitudes for $S\geq 1$ are
\begin{eqnarray} 
\label{Vac} 
&& V_{q}^{cc}(k,k')=V_{q}^{ca}(k,k')=J_{q}-\frac{1}{2}(J_{k}+J_{k'}),\nonumber\\
&& V_{q}^{ac}(k,k')=J_{q}-J_{\perp}-\frac{1}{2}(J_{k}+J_{k'}), \\
&& V_{q}^{aa}(k,k')=J_{q}+J_{\perp}-\frac{1}{2}(J_{k}+J_{k'}),\nonumber
\end{eqnarray}
and for $S=\frac{1}{2}$  they have to be modified as
\begin{eqnarray} 
\label{Vac-s12} 
&& V_{q}^{cc}=V_{q}^{aa}=\frac{1}{2}(U+J_{\perp})+J_{q},\nonumber\\
&& V_{q}^{ac}= V_{q}^{ca}=\frac{1}{2}(U-J_{\perp})+J_{q}.
\end{eqnarray}

\begin{figure}[tb]
 \includegraphics[width=0.42\textwidth]{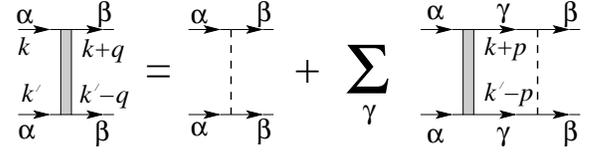}
 \caption{
\label{fig:Bethe-Salpeter} 
The ladder approximation to the Bethe-Salpeter equation for the
renormalized two-body interaction vertex
$\Gamma_{q}^{\beta\alpha}(k,k';E)$. Solid lines denote bare
propagators. The approximation becomes exact at $\mu\to 0$.}
\end{figure}

We have thus mapped the initial spin problem onto a 1d lattice gas of
particles with a nontrivial double-minima dispersion.  The
renormalized two-body interaction in such a gas can be easily found in
the dilute limit, i.e., $\mu\to 0$. Since the self-energy vanishes at
 $\mu\to 0$, the full propagator coincides with the bare one \cite{Uzunov81}, and thus the Bethe-Salpeter (BS) equation for
the renormalized two-body interaction vertex $\Gamma_{q}^{\alpha\beta}
(k,k';E)$ (where $E$ is the total energy of the incoming particles)  takes the following 
form:\cite{BatyevBraginskii84}
\begin{eqnarray} 
\label{BS-gen}
&& \Gamma_{q}^{\alpha\beta}(k,k';E)=V_{q}^{\alpha\beta}(k,k')\\
&& \qquad -\frac{1}{L}\sum_{p}\sum_{\gamma}\frac{V_{q-p}^{\alpha\gamma}(k+p,k'-p)\Gamma_{p}^{\gamma\beta}(k,k';E)}
{\varepsilon_{k+p}^{\gamma}+\varepsilon_{k'-p}^{\gamma}-E},\nonumber
\end{eqnarray}
where labels $\alpha$, $\beta$, $\gamma$ denote the magnon branch and can
take the values ``$a$'' and ``$c$''.
The above equation is schematically shown in terms of Feynman diagrams in Fig.~(\ref{fig:Bethe-Salpeter}).

If the magnetic field is close enough to the saturation, so that the
condition
\begin{equation} 
\label{vicinity-2} 
\mu=H_{s}-H \ll \Delta_{2}\equiv 2J_{2}S(|\beta|/4-1)^{2}
\end{equation} 
is satisfied (see Fig.\ \ref{fig:ac-magnons}), then the system
is mainly populated by magnons (of $a$- or $c$-branch, depending on
the sign of $J_{\perp}$) with momenta around the two
dispersion minima at $\pm Q$, which at low energies can be interpreted
as two different bosonic ``flavors''. For those low-energy modes, one
can formulate  the effective theory in the form of the
 Gross-Pitaevsky-type energy functional for a two-component Bose field:
\begin{eqnarray} 
\label{GP} 
\mathcal{H}_{\rm GP}&=&\int dx\,\Big\{ \sum_{\sigma=1,2}\frac{|\nabla
\Phi^{\sigma}|^{2}}{2m}
+\frac{1}{2}\Gamma_{11}(n_{1}^{2}+n_{2}^{2})\nonumber\\
&+&\Gamma_{12} n_{1}n_{2} -\mu(n_{1}+n_{2})\Big\}.
\end{eqnarray}
Here the Planck constant is set to unity, $\Phi^{1,2}$ are the   macroscopic 
bosonic fields that describe magnons with momenta $k$ lying within the
intervals $|k\pm Q|<\Lambda$ around the dispersion minima, $\Lambda$
is the  infrared cutoff,  $n_{\sigma}=|\Phi^{\sigma}|^{2}$ are the
particle densities,  and $m$ is the effective mass,
\begin{equation} 
\label{mass} 
\frac{1}{m}=\left.\frac{\partial^{2}\varepsilon_{k}}{\partial k^{2}}\right|_{k=Q} =\frac{SJ_2 (16-\beta^{2})}{2}.
\end{equation} 
The point $\Gamma_{11}=\Gamma_{12}$
corresponds to the enhanced $SU(2)$ symmetry at the level of the
effective low-energy theory.
For $\Gamma_{12}<\Gamma_{11}$,  the ground state of the gas contains an equal
density of the two particle species, and for  $\Gamma_{12}>\Gamma_{11}$
 just one
of the two species is present in the ground state. 
In the spin problem the total number of each
bosonic species is not separately fixed, in contrast to a typical setup for
atomic mixtures. In a setup with fixed particle numbers, the
ground state
at  $\Gamma_{12}<\Gamma_{11}$ is in the mixed phase, and
$\Gamma_{12}>\Gamma_{11}$ corresponds to phase separation. 
In the spin language, the separated phase maps to the vector chiral (VC)
phase \cite{KV05}, while the mixed phase corresponds
to the  two-component Tomonaga-Luttinger 
liquid (TLL2) \cite{FathLittlewood98,Okunishi+99}.

Macroscopic effective couplings $\Gamma_{11}$, $\Gamma_{12}$ of the
Gross-Pitaevsky-type theory can be obtained from the $E\to 0$ limit of
the corresponding vertex functions:
\begin{eqnarray} 
\label{Gamma-Lambda} 
\Gamma_{11}&=& \left.\Gamma_{0}^{\alpha\alpha}(Q,Q;0)\right|_{\Lambda=\Lambda_{*}},\nonumber\\
\Gamma_{12}&=& \left.\big[ \Gamma_{0}^{\alpha\alpha}(-Q,Q;0) + 
\Gamma_{2Q}^{\alpha\alpha}(-Q,Q;0)\big]\right|_{\Lambda=\Lambda_{*}},
\end{eqnarray}
where $\alpha=$``$a$'' or ``$c$'' for positive and negative
$J_{\perp}$, respectively, and the vertex function
$\Gamma_{q}^{\alpha\alpha}(k,k';E)$ is the solution of the BS equation
(\ref{BS-gen}) with the infrared cutoff $|p|>\Lambda$ employed in
the summation over internal transferred momenta $p$.  The resulting
expressions are viewed as functions of the running cutoff $\Lambda$ in
the spirit of the renormalization group (RG) approach, and the RG flow
$\Lambda\to 0$ is then interrupted at a certain scale
$\Lambda=\Lambda_{*}=\sqrt{\mu m/2}$ that depends on the chemical
potential (or, in other words, on the magnon density).  The above
approach is well known for one-component Bose gas
\cite{FisherHohenberg88,NelsonSeung89,KolomeiskyStraley92,Kolomeisky+00}
and has been successfully applied to the multicomponent case recently
\cite{Kolezhuk10,Kolezhuk+12}.

There is an alternative approach \cite{Lee+02,Morgan+02,Kolezhuk+12}, which,
instead of the infrared cutoff in the momentum space, introduces an
``off-shell'' regularization: the two-body scattering amplitudes in
the presence of a finite  particle density are obtained by taking the
``bare'' expressions  $\Gamma_{q}^{\alpha\beta}(k,k';E)$ at a finite
\emph{negative} energy $E=-E_{*}=-\pi^{2}\mu/8$:
\begin{eqnarray} 
\label{Gamma-E} 
\Gamma_{11}&=& \left.\Gamma_{0}^{\alpha\alpha}(Q,Q;-E_{*})\right|_{\Lambda=0},\\
\Gamma_{12}&=& \left.\big[
  \Gamma_{0}^{\alpha\alpha}(-Q,Q;-E_{*})+\Gamma_{2Q}^{\alpha\alpha}(-Q,Q;-E_{*}) \big]\right|_{\Lambda=0},\nonumber
\end{eqnarray}
where $\alpha$ takes the same value as in
Eq.\ (\ref{Gamma-Lambda}). One can show \cite{Kolezhuk+12} that the off-shell
regularization yields the results that are equivalent to the cutoff
regularization scheme. In this work, we have used the off-shell
regularization because it is more convenient technically.

Our model contains only short-range interactions, so the solution of
(\ref{BS-gen}) can be expressed in terms of a finite number of Fourier
modes in the transferred momentum \cite{BatyevBraginskii84}. In our
case, from the structure of $V_{q}^{\alpha\beta}(k,k')$ it is easy to
see that each component of $\Gamma_{q}^{\alpha\beta}(k,k';E)$ can contain
only five Fourier harmonics proportional to $1$, $\cos q$, $\sin q$,
$\cos 2q$, and $\sin 2q$.  The system of integral equations is thus
reduced to a system of linear equations that can be solved
analytically for any value of the spin $S$.

For the purpose of finding only the Gross-Pitaevsky  effective couplings $\Gamma_{11}$,
$\Gamma_{12}$, the problem can be simplified even further. First of
all, the system (\ref{BS-gen}) describes four equations for the
vertices, which split into two decoupled pairs: a pair of coupled
equations for $\Gamma^{aa}$, $\Gamma^{ca}$, and another pair of
coupled equations for $\Gamma^{cc}$ and $\Gamma^{ac}$.  For
$J_{\perp}>0$, the lowest energy excitations are $a$-magnons, and
thus, in order to find the effective couplings $\Gamma_{11}$,
$\Gamma_{12}$, we only need to solve the first pair of the BS
equations for $\Gamma^{aa}$, $\Gamma^{ca}$; similarly, for
$J_{\perp}<0$ we are interested only in the equations for
$\Gamma^{cc}$ and $\Gamma^{ac}$.
Second,  it is easy to see that $\Gamma_{11}$ can be found as the $q=0$
value of $\Gamma^{\alpha\alpha}_{q}(Q,Q;-E_{*})$ which is
an even function of the transferred momentum $q$, and $\Gamma_{12}$
may be represented as the $q=Q$ value of the function
$\Gamma^{\alpha\alpha}_{Q+q}(-Q,Q;-E_{*})+\Gamma^{\alpha\alpha}_{Q-q}(-Q,Q;-E_{*})$,
which is also even in $q$. For that reason, one can rewrite the integral
equations (\ref{BS-gen}) for the above two even functions,  keeping
only even Fourier harmonics \cite{UedaTotsuka09}. This reduces the number of resulting linear equations
to six and makes the problem amenable to analytical treatment. We refer
the reader to the Appendix for further details.

Solving the Bethe-Salpeter equation, one can show that the expansion
of $\Gamma_{11}$, $\Gamma_{12}$ in $E_{*}$ has the following
structure:\cite{Kolezhuk+12}
\begin{eqnarray} 
\label{enG-series} 
\frac{1}{\Gamma_{11}}&=&\left(\frac{m}{4E_{*}}\right)^{1/2} +
\frac{1}{g_{11}}+O(E_{*}^{1/2})+\ldots,\nonumber\\
\frac{1}{\Gamma_{12}}&=&\left(\frac{m}{4E_{*}}\right)^{1/2} +
\frac{1}{g_{12}}+O(E_{*}^{1/2})+\ldots
\end{eqnarray}
Note that 
for $E_{*}\to0$ (i.e., $\mu\to 0$) the effective couplings $\Gamma_{11}$ and $\Gamma_{12}$
flow to the same  value, which reflects the tendency of the
RG flow to restore the $SU(2)$ symmetry for the two-component Bose
mixture.\cite{Kolezhuk10} 

Parameters $g_{11}$, $g_{12}$, entering the second term in the expansions
(\ref{enG-series}), 
under certain conditions, namely $|g_{11}|m \ll 1$ and $|g_{12}|m \ll 1$,
can be identified with
the effective bare coupling constants of the continuum two-component Bose gas with
contact interactions (see Ref.\ \onlinecite{Kolezhuk+12} for details).
If the above conditions  are broken,
parameters $g_{ij}$ cannot be interpreted as physical bare
couplings, and only the  renormalized interactions $\Gamma_{ij}$
retain their meaning as
effective low-energy coupling constants. 
The only physical meaning
of $g_{ij}$ in such a case is that they are connected to the
asymptotic phase shift of scattering states at  small transferred
momenta \cite{Kolezhuk+12}.

From Eqs.\ (\ref{enG-series}) one can see that transition points between
the TLL2 and VC phases, that are determined by the condition $\Gamma_{11}=\Gamma_{12}$,
correspond only to  
\emph{crossings} of $g_{12}$ and
$g_{11}$. (See, for example, Fig.\ \ref{fig:g-S1}a: when $g_{11}$ goes through a pole changing sign
from plus to minus infinity, then at the pole
$g_{12}-g_{11}$ changes sign from negative
to positive, but this does not correspond to any phase
transition, because on both sides of the pole in its immediate
vicinity $\Gamma_{12}<\Gamma_{11}$). 

Similarly,  $g_{11}$ or $g_{12}$ becoming
negative by going through a \emph{zero} indicates the
appearance of magnon bound states with zero total momentum, while a
change of sign through a \emph{pole} is not signaling any transition,
but rather indicates a crossover into the so-called ``super-Tonks''
regime \cite{Astrakharchik-superTonks,note-superTonks}.
It should be emphasized that within the present effective
theory, which is essentially based on the two-body interaction, we
cannot predict whether the formation of bound states stops at the
level of bound magnon pairs, or continues with multiparticle bound
states.

\section{Strong-field phase diagram: analytical  results }
\label{sec:tube-results}

Let us turn our attention to the specific
predictions of our theory for the frustrated spin tube model
defined by the Hamiltonian (\ref{hamS}), at two spin values  $S=1$ and
$S=\frac{1}{2}$.  Details concerning the solution of the BS equations
can be found in the Appendix.

\begin{figure*}[tb]
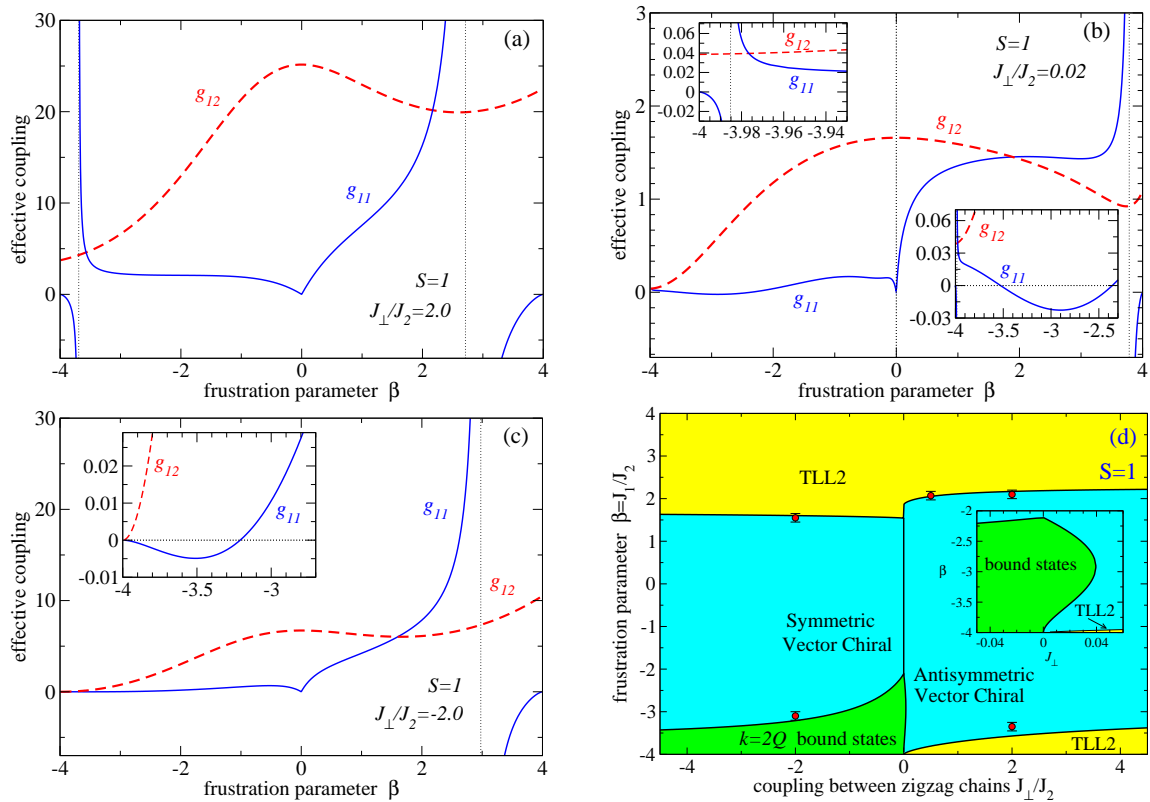

 \includegraphics[width=0.4\textwidth]{g-S1-J2.00.eps}
\hspace*{7mm}\includegraphics[width=0.4\textwidth]{g-S1-J0.02.eps}

\includegraphics[width=0.4\textwidth]{g-S1-J-2.00.eps}
\hspace*{7mm}\includegraphics[width=0.4\textwidth]{s1-double-phd.eps}
 \caption{
\label{fig:g-S1} (Color online). (a)
``Bare'' coupling constants $g_{11}$,
$g_{12}$  of the effective theory,  obtained from the analytical solution of 
regularized Bethe-Salpeter equations, for $S=1$ frustrated spin
tube with  $J_{\perp}/J_{2}=2$; (b), (c) the same for  $J_{\perp}/J_{2}=0.02$ and
 $J_{\perp}/J_{2}=-2$; (d) the predicted phase diagram in the vicinity
of saturation; symbols show the transition points obtained from
numerical simulations. Crossings of $g_{11}$ and $g_{12}$
correspond to transitions between the vector chiral and two-component
Luttinger liquid phases, while the regions where $g_{11}$ becomes
negative by going through a \emph{zero} (not through a \emph{pole}) indicate the
appearance of magnon bound states. }
\end{figure*}

Fig.\ \ref{fig:g-S1}a-c illustrates  the behavior of
the ``bare'' effective couplings $g_{11}$, $g_{12}$ for the spin-$1$
tube, as functions of
the frustration parameter $\beta$, for several values of the interchain
coupling $J_{\perp}$.  VC-TLL2 transitions are detected by crossings
of $g_{11}$ and $g_{12}$, and zeros of $g_{11}$ signal the formation
of magnon bound states with the total momentum $k=\pm 2Q$. The resulting phase diagram is shown in Fig.\ \ref{fig:g-S1}d.
 One can see that the region of small $|\beta|$ 
always corresponds to the chiral phase, similar to the case of a
single frustrated spin-$1$ chain \cite{Arlego+11,Kolezhuk+12}. For
antiferromagnetic zigzag coupling $\beta>0$, there is only one
transition between the vector chiral and the two-component Luttinger
liquid phases, which is rather weakly dependent on the interchain
(rung) coupling $J_{\perp}$. Nonanalytic behavior of the phase boundary at
$J_{\perp}=0$ stems from the following: we assume that we work in the immediate
vicinity of the saturation field (see conditions (\ref{vicinity}) and
(\ref{vicinity-2})), the ground state contains only $a$-magnons at
$J_{\perp}>0$ and only $c$-magnons 
at $J_{\perp}<0$. It is clear that the magnetic field range, where
the assumption (\ref{vicinity}) remains applicable, shrinks to zero as $J_{\perp}\to
0$, which causes the above nonanalyticity. At $J_{\perp}=0$ the model corresponds to two decoupled frustrated
chains, so two magnon branches become degenerate, and the problem
reduces to that for a single chain \cite{Kolezhuk+12}.

For ferromagnetic zigzag coupling $\beta<0$, there is a large region
with negative $g_{11}$ supporting  bound magnon states.
 From the numerical analysis for a single $S=1$ frustrated
chain \cite{Arlego+11}, it is known that  at least around
$J_{\perp}=0$ there is a metamagnetic jump in the magnetization curves
in this region. Presence of such a jump indicates formation of
``magnon drops'' -- bound states of a large number of magnons.
Increasing antiferromagnetic rung interaction $J_{\perp}>0$ leads to
the opening of a finite TLL2 phase window close to $\beta=-4$
(which is the boundary of a transition into a one-component Luttinger
liquid state).

\begin{figure}[tb]
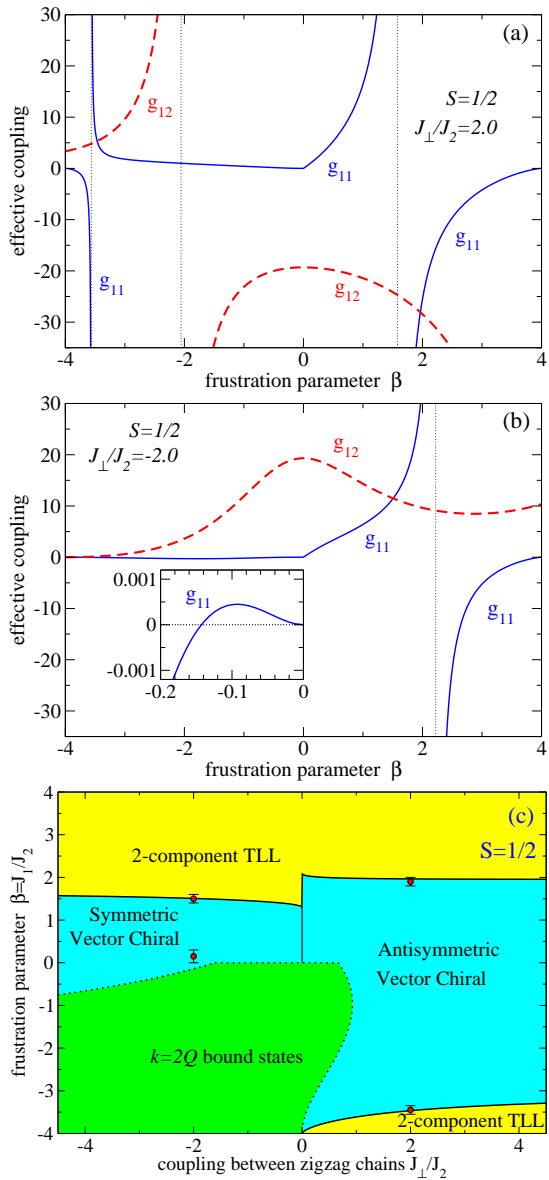

 \includegraphics[width=0.4\textwidth]{g-S12-J2.00.eps}

\includegraphics[width=0.4\textwidth]{g-S12-J-2.00.eps}

\includegraphics[width=0.4\textwidth]{s12-double-phd.eps}
 \caption{
\label{fig:g-S12}  (Color online).
``Bare'' coupling constants $g_{11}$,
$g_{12}$  for $S=\frac{1}{2}$ frustrated spin
tube with (a) $J_{\perp}/J_{2}=2$ and (b) $J_{\perp}/J_{2}=-2$; (c) the predicted phase diagram in the vicinity
of saturation, symbols show the transition points obtained from
numerical simulations. }
\end{figure}

Fig.\ \ref{fig:g-S12} shows two examples of the characteristic
behavior of the bare coupling, along with the resulting phase diagram,
for the $S=\frac{1}{2}$ tube. The topology of the phase diagram is
qualitatively the same as in the $S=1$ case, but there are certain
caveats which one should have in mind. One important difference
concerns the region of bound magnon states. It is known
\cite{KuzianDrechsler07} that for a single isotropic
$S=\frac{1}{2}$ frustrated chain with $\beta<0$ the lowest energy of
a two-magnon bound state is not reached at the total momentum $k=\pm
2Q$, as one could expect from the picture of two bound magnons of the
same flavor, but instead 
the minimum of the bound state dispersion lies at
$k=\pi$ in a rather wide region of $-2.67<\beta< 0$. For that reason, one may
expect that the actual size of the region dominated by magnon bound
states is  larger than that of the region labeled ``$k=2Q$ bound
states'' in Fig.\ \ref{fig:g-S12}. Second, from the analysis of the
single-chain $S=\frac{1}{2}$ problem in
Ref.\ \onlinecite{Kolezhuk+12}, it is known that predictions of the
present theory  for the VC-TLL2 transition at $J_{\perp}=0$ should deviate substantially from
the numerical results. In Section \ref{sec:numerics}
we will see, however, that agreement with the numerics is actually
improved with the increase of the rung coupling strength $|J_{\perp}|$.

\section{Numerical analysis}
\label{sec:numerics}

To verify our analytical predictions, we have studied the frustrated
spin tube model (\ref{hamS}) with $S=1$ and $S=\frac{1}{2}$ 
using the
density matrix renormalization group \cite{white92} (DMRG) 
method 
(see  Ref.\ \onlinecite{schollwoeck05} for
a detailed description of the DMRG technique). 

We 
study the vector chirality correlation functions to identify the phases that
have long-range vector chiral order. The ground state of spin tubes with $J_{\perp}>0$
near the saturation is populated with the magnons of
the antisymmetric $a$-branch, so
the relevant quantity in that case is the \emph{antisymmetric chirality}
\begin{equation} 
\label{kappa-A} 
\vec{\kappa}_{A}(n)=(\vec{S}_{n,1}-\vec{S}_{n,2})\times (\vec{S}_{n+1,1}-\vec{S}_{n+1,2}).
\end{equation}
Similarly, for $J_{\perp}<0$ one has to look at the correlators of the \emph{symmetric chirality}
\begin{equation} 
\label{kappa-S} 
\vec{\kappa}_{S}(n)=(\vec{S}_{n,1}+\vec{S}_{n,2})\times (\vec{S}_{n+1,1}+\vec{S}_{n+1,2}).
\end{equation}

In order to identify regions with metamagnetic behavior, i.e., the
regions where magnon attraction leads to the formation of a single
bound state consisting of a macroscopic number of magnons (``magnon
drop''), we calculate the distribution of the rung magnetization
$M_{n}=\langle S_{n,1}^{z}+ S_{n,2}^{z}\rangle$ along the tube.  This
approach, however, does not allow to detect phases where the formation
of bound states stops at the level of a finite number of magnons.

We use DMRG in its matrix product state formulation
\cite{McCullochGulacsi02,McCulloch07}, which allows us to exploit the
non-Abelian $SU(2)$ symmetry, as well as the Abelian $U(1)$.  (While
the magnetic field $H$ breaks $SU(2)$ symmetry, the fact that the
Zeeman energy term commutes with the rest of the Hamiltonian makes it
possible to take the influence of the magnetic field into account by
calculating the ground state of the model in a sector with the given
total spin $S_{\rm tot}$.) The advantage of using the $SU(2)$ symmetry
lies in a considerable reduction of the number of states $m$ which is
necessary to describe the system, because one essentially treats the
multiplet of states of the same total spin as a single representative
state.

The use of $SU(2)$ symmetry has a disadvantage as well: since the
non-Abelian method allows to compute only reduced matrix elements (in
the sense of the Wigner-Eckart theorem), one can only compute
rotationally invariant correlators such as $\langle \vec{\kappa}_{A}(n)\cdot
\vec{\kappa}_{A}(n')\rangle$, etc. This can be inconvenient if the
contribution of the transversal components of chirality exhibits
strong oscillations that act as a ``noise'' masking the long-range
order in the longitudinal component, as it has been found in
frustrated chains \cite{McCulloch+08}. We have found such strong
oscillations for $\vec{\kappa}_{S}$ in spin tubes with $J_{\perp}<0$,
while the correlators of $\vec{\kappa}_{A}$ in $J_{\perp}>0$
systems were essentially free from oscillations.  
calculating 

\begin{figure}[tb]
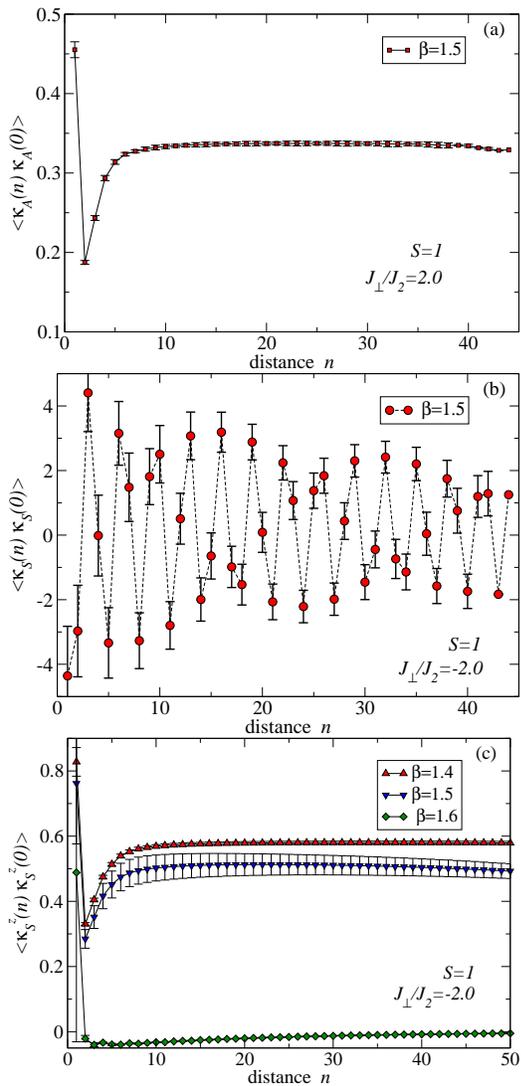

\includegraphics[width=0.38\textwidth]{Fkappa-S1a.eps}

\includegraphics[width=0.38\textwidth]{Fkappa-S1b.eps}

\includegraphics[width=0.38\textwidth]{Fkappa-S1c.eps}

\caption{\label{fig:correls}  (Color online).
Typical DMRG results for the chirality correlators of the 
$L=64$ spin-1 tube ($128$ spins) at the total
magnetization $S_{\rm tot}=116$, inside the vector chiral phase:
(a) rotationally invariant correlator of the antisymmetric chirality $\langle\vec{\kappa}_{A}(n)\cdot
\vec{\kappa}_{A}(0)\rangle$, calculated with the use of the $SU(2)$ symmetry: there are no visible oscillations; 
(b)  rotationally invariant correlator of the symmetric chirality $\langle\vec{\kappa}_{S}(n)\cdot
\vec{\kappa}_{S}(0)\rangle$ exhibits strong oscillations which are  due to the
contribution of the transversal components as seen from (c)  longitudinal
correlators of the same quantity.
}
\end{figure}

For that
reason, we have used $SU(2)$ symmetry in our calculations for
antiferromagnetically  coupled tubes ($J_{\perp}>0$), and resorted to
the standard $U(1)$ calculations for the  $J_{\perp}<0$ case.
Fig.\ \ref{fig:correls} shows  typical examples of chiral
correlators for systems with ferro- and antiferromagnetic sign of
$J_{\perp}$, calculated with or without the use of the $SU(2)$ symmetry.

We have studied spin-$1$ and spin-$\frac{1}{2}$ spin tubes consisting
of up to $256$ spins, with open boundary conditions, keeping up to
$m=600$ states in most calculations. The total magnetization
$M$ of the system has been set at about $90\%$ of the saturation value $M_{s}$
(specifically, we have kept $M/M_{s}=115/128$ for $S=1$ tubes and
$M/M_{s}=29/32$ for $S=\frac{1}{2}$ ones).
The (squared) chiral order parameters $\kappa_{A}^{2}$,
$\kappa_{S}^{2}$ were extracted from the large-distance behavior of the corresponding
correlation functions (the technicalities of this procedure are described in detail in Ref.\ \onlinecite{Mc+08}). 

\begin{figure}[tb]
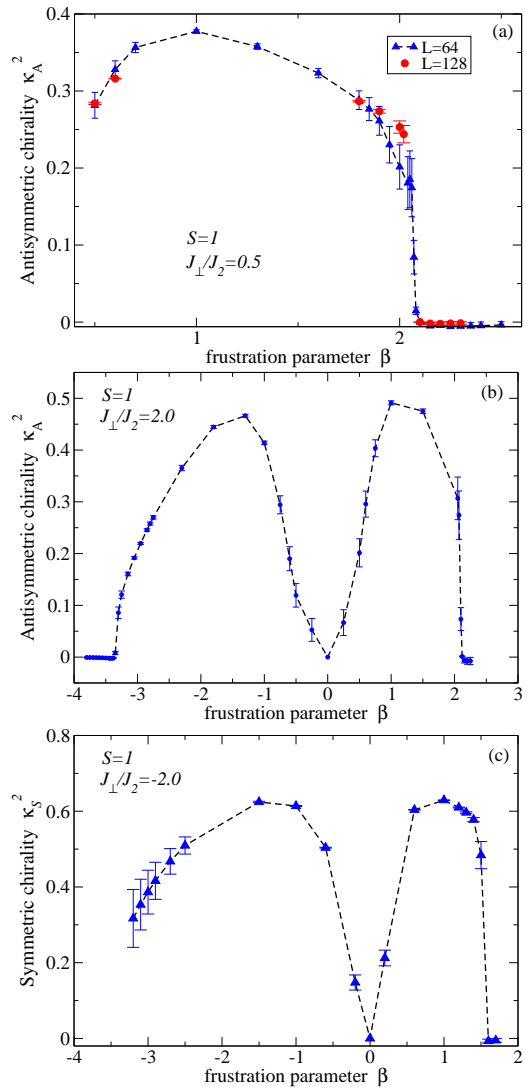

\includegraphics[width=0.38\textwidth]{kappaA-S1-J0.50.eps}

\includegraphics[width=0.38\textwidth]{kappaA-S1-J2.00.eps}

\includegraphics[width=0.38\textwidth]{kappaS-S1-J-2.00.eps}

\caption{
\label{fig:kappa-S1}  
(Color online).
Chirality order parameters of the $S=1$ tube obtained from the
large-distance behavior of the correlation functions. The results for
$L=64$ tube ($128$ spins) were calculated in the sector with the total
spin $S_{\rm tot}=116$, and for $L=128$ we took  $S_{\rm
  tot}=230$.  
(a) $J_{\perp}/J_{2}=0.5$; 
(b)  $J_{\perp}/J_{2}=2.0$, only $L=64$ results are shown;
(c)  $J_{\perp}/J_{2}=-2.0$, here the calculations have been done using
the usual $U(1)$ DMRG method for $L=128$ system.
The point at $\beta=0$ has not been obtained numerically, but is
included as a guide to the eye since the chirality must vanish at $\beta=0$.
}
\end{figure}

Fig.\ \ref{fig:kappa-S1} shows the behavior of
the chiral order parameters along three constant-$J_{\perp}$ cuts in
the phase diagram of the spin-$1$ tube.  For antiferromagnetic rungs
($J_{\perp}/J_{2}=2$) one can clearly see two transitions around
$\beta\approx 2.1$ and $\beta\approx -3.40$, which is consistent with
the predictions of our theory (see Fig.\ \ref{fig:g-S1}d).  At both
transitions $\kappa_{A}$ vanishes in a rather abrupt manner, which
suggests that the transition is of the first order.
One can also notice that the amplitude of the chiral order decreases
as $\beta$ tends to $0$, which is explained by the fact that at
$\beta=0$ the chirality should vanish since this limit
corresponds to two decoupled unfrustrated ladders.

\begin{figure}[tb]
\includegraphics[width=0.42\textwidth]{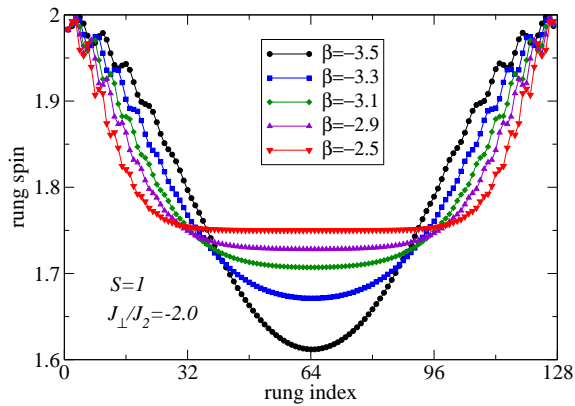}
\caption{
\label{fig:Srung-S1a}  
(Color online).
The distribution of the rung magnetization $M_{n}=\langle S_{n,1}^{z}+
S_{n,2}^{z}\rangle$ along the $S=1$ spin tube of the length $L=128$,
at fixed $J_{\perp}/J_{2}=-2$ and several different values of the
frustration parameter $\beta=J_{1}/J_{2}$ in the vicinity of
the transition from the repulsive to the attractive magnon gas.
}
\end{figure}

\begin{figure}[tb]
\includegraphics[width=0.42\textwidth]{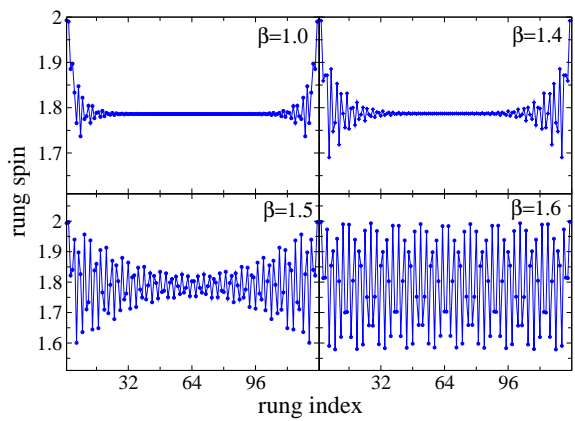}
\caption{
\label{fig:Srung-S1b}  (Color online). The same as in
Fig.\ \ref{fig:Srung-S1a}, in the vicinity of the transition from the
vector chiral into the two-component Tomonaga-Luttinger liquid phase.
}
\end{figure}

For ferromagnetic rungs, 
$J_{\perp}/J_{2}=-2$, there is a transition in $\kappa_{S}$ at positive
$\beta\approx 1.6$ which has a similar
behavior to the corresponding transition in $\kappa_{A}$ at
$J_{\perp}>0$, while the situation at negative $\beta$ is
different: Looking simply at the chiral correlation functions, one
tends to think that the chirality persists all the way up to
$\beta=-4$, and merely the influence of the boundaries seems to
increase considerably at $\beta\lesssim -3.0$.
Fig.\ \ref{fig:Srung-S1a} shows how the distribution of the
magnetization along the tube changes in this region of $\beta$.
One can see that there is a transition
at around $\beta\approx -3.2$ which corresponds to the formation of a
macroscopic magnon ``droplet'' in the ground state, sitting in the
middle of the system. This is exactly  the behavior found in
Ref.\ \onlinecite{Arlego+11} for ferromagnetic frustrated $S=1$
chains, and indicates that the region with $\beta\lesssim -3.2$
exhibits a \emph{metamagnetic jump} in the magnetization curve (the states
with a ``droplet'' are never realized as true ground states at fixed 
magnetic field, they are only possible if the number of magnons is
artificially fixed). 

It is worthwhile to remark that the behavior of
the magnetization distribution can be also used to detect the
transition between the VC and TLL2 phases. Fig.\ \ref{fig:Srung-S1b}
shows how the magnetization oscillations, which are localized at the
boundaries in the VC phase, penetrate the bulk  and spread
over the entire system when one moves across the point
$\beta\approx 1.5$. Comparing this behavior with Fig.\ \ref{fig:kappa-S1}c, one
can see that  $\beta\approx 1.5$ is indeed the transition point
where the symmetric chirality vanishes.

\begin{figure}[tb]
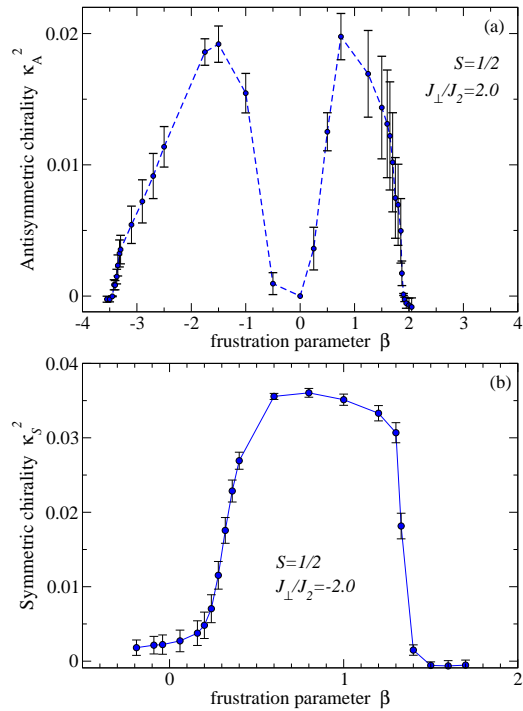

\includegraphics[width=0.38\textwidth]{kappaA-S12-J2.00.eps}

\includegraphics[width=0.38\textwidth]{kappaS-S12-J-2.00.eps}
\caption{
\label{fig:kappa-S12}  
(Color online).
Chirality order parameters of the $S=\frac{1}{2}$ tube extracted from the
large-distance behavior of the correlation functions: 
(a) shows results
obtained by means of the $SU(2)$-symmetric DMRG method for 
$L=64$ system ($128$ spins) with  $J_{\perp}/J_{2}=2.0$, in the sector with the total
spin $S_{\rm tot}=116$; the point at $\beta=0$ is included as a guide
to the eye; (b) the results obtained by the standard $U(1)$-symmetric DMRG
calculation for $L=128$ tube with  $S_{\rm  tot}=230$ and  $J_{\perp}/J_{2}=-2.0$. 
}
\end{figure}

\begin{figure}[tb]
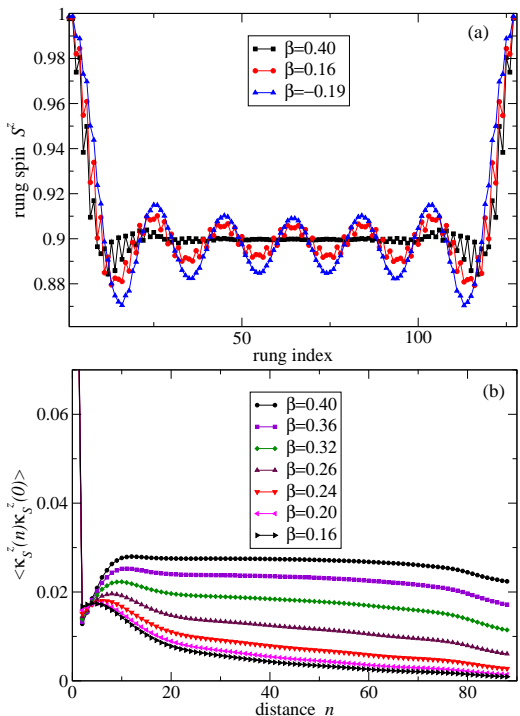

\includegraphics[width=0.38\textwidth]{Srung-S12-J-2.00.eps}

\includegraphics[width=0.38\textwidth]{kappa-S12-J-2.00.eps}
\caption{
\label{fig:Srung-S12}  (Color online). (a) The distribution of the rung magnetization $M_{n}=\langle S_{n,1}^{z}+
S_{n,2}^{z}\rangle$ along the  $L=128$ spin-$\frac{1}{2}$ tube,
at fixed $J_{\perp}/J_{2}=-2$ and several values of the
frustration parameter $\beta$ in the vicinity of the transition
from the vector chiral phase to the phase with bound magnons and
dominating spin density wave correlations.
}
\end{figure}

For $S=\frac{1}{2}$ tube, we have done similar calculations as for the
spin-$1$ case, along two cuts at $J_{\perp}=\pm 2J_{2}$  in the phase diagram. The resulting
behavior of the chiral order parameters along those lines is shown in
Fig.\ \ref{fig:kappa-S12}. While at $J_{\perp}>0$ the picture is
essentially similar to that for spin-$1$ system, as described above, at
negative $J_{\perp}$ the transition at lower values of $\beta$ looks
rather different: the chiral order parameter $\kappa_{S}$ disappears
in a very smooth way, as seen from Fig.\ \ref{fig:kappa-S12}b (the
corresponding correlation functions of the symmetric chirality are
presented in Fig.\ \ref{fig:Srung-S12}b). At the same time, the
distribution of magnetization at this transition shows the development
of a spin density wave as seen in Fig.\ \ref{fig:Srung-S12}a. This
is reminiscent of  what happens in ferromagnetic frustrated
spin chains \cite{Hikihara+08,Sudan+09,HM+09}, and indicates that this transition
corresponds to the formation of bound states of finite number of
magnons, in contrast to the $S=1$ case where there is a single
bound state absorbing all the magnons present in the
system.  

Comparing our numerical results for the selected cuts in the phase
diagram with the analytical predictions of the  two-component Bose gas
approach, one can see that our theory captures fairly well the physics
of phase transitions in the frustrated spin tube model (\ref{hamS}). 
Comparing the results for the  $S=\frac{1}{2}$ spin tube (double zigzag
chain) with those obtained by the same approach for single
$S=\frac{1}{2}$ zigzag chain \cite{Kolezhuk+12}, one can see that the 
accuracy of the prediction for the VC-TLL2 transition is improved
when one includes sufficiently strong rung coupling $|J_{\perp}|$.
One obvious drawback of the theory, as mentioned at the end of 
Sect.\ \ref{sec:tube-results}, is that it does not
take into account the possibility to have the lowest energy of the
bound state at the total momentum different from $\pm 2Q$, which is
realized for $S=\frac{1}{2}$. We see that
for that
reason, our analytical predictions underestimate the size of the
region dominated by bound states for $S=\frac{1}{2}$. Apart from that,
one may call the agreement between the analytical theory and numerical
simulations satisfactory.

\section{Summary}
\label{sec:summary}

We study the ground state phase diagram of  strongly frustrated
four-leg spin-$S$ tube (which may be alternatively represented as two coupled zigzag
spin chains, or as a two-leg spin ladder with 
next-nearest-neighbor couplings along the legs, see
Fig.\ \ref{fig:tube}) in a strong magnetic field in the 
vicinity of saturation. The model  is motivated by the physics
of such frustrated quasi-one-dimensional spin-$\frac{1}{2}$ materials as Sulfolane-$\rm
Cu_{2}Cl_{4}$ \cite{Fujisawa+03,Garlea+08,Garlea+09,Zheludev+09} and
$\rm
BiCu_{2}PO_{6}$\cite{Koteswararao+07,Mentre+09,Tsirlin+10,Lavarelo+11},
but is also interesting in itself as the simplest model of coupled
frustrated chains.
Although both in Sul-$\rm Cu_{2}Cl_{4}$ and $\rm BiCu_{2}PO_{6}$ the
saturation field is too high to be accessible in current experiments,
we hope that our findings, which establish the high-field slice  of the
phase diagram, will stimulate experimental studies of
field-induced  phases in those systems. 

In the vicinity of saturation, the system can be represented as a
dilute gas of two flavors of bosonic quasiparticles corresponding to
magnons with momenta around two degenerate incommensurate minima of
the magnon dispersion.  Using the method previously proposed for
frustrated spin chains \cite{Arlego+11,Kolezhuk+12}, we calculate
effective interactions in this two-component Bose gas, and establish
the high-field phase diagram of the frustrated spin tube.  We show
that the phase diagram contains two types of vector chiral phases
(with symmetric and antisymmetric long-range chiral order), the
two-component Luttinger liquid, and phases dominated by the presence
of bound magnons.
 
We complement our analytical results by the numerical studies of $S=1$ and $S=\frac{1}{2}$
frustrated tubes by means of the
density matrix renormalization group technique. We analyze the
behavior of chiral correlation functions and
distribution of the magnetization along several cuts in the phase
diagram, and extract the position of the corresponding phase boundaries.
The numerical results are found to be consistent with our analytical predictions.

\begin{acknowledgments}

We thank G. Roux and T. Vekua for useful discussions.  A.K. gratefully
acknowledges the hospitality of the Laboratoire de Physique
Th\'eorique et Mod\'eles Statistiques at Universit\'e Paris Sud, and
of the Institute for Theoretical Physics at the Leibniz University of
Hannover during research stays that have led to the initiation of this study.  
This work has been partly supported by the State Program
``Nanotechnologies and Nanomaterials'' of the Government of Ukraine,
Project 1.1.3.27, and by the Program 11BF07-02 from the Ministry
of Education of Ukraine. Numerical calculations have been performed on
the computing cluster of V. E. Lashkarev Institute of Semiconductor
Physics.

\end{acknowledgments}

\appendix*

\section{\boldmath Computing the Gross-Pitaevsky  couplings  $\Gamma_{11}$, $\Gamma_{12}$}
\label{app:BS-sym}

Consider in some detail the procedure of solving the Bethe-Salpeter
equations (\ref{BS-gen}). To reduce the number of Fourier harmonics, 
we first symmetrize the kernel. In doing so, we use the identities
$\varepsilon_{k}^{\alpha}=\varepsilon_{-k}^{\alpha}$, $V_{q}^{\alpha\beta}(k,k')=V_{-q}^{\alpha\beta}(k',k)$, and
$\Gamma_{q}^{\alpha\beta}(k,k';E)=\Gamma_{-q}^{\alpha\beta}(k',k;E)$.
 Let us introduce the following
functions that are even in the transferred momentum $q$:
\begin{eqnarray} 
\label{sym-Gamma}
&& A_{q}^{\alpha\beta} \equiv \Gamma_{q}^{\alpha\beta}(Q,Q;-E_{*}),  \\
&& B_{q}^{\alpha\beta}\equiv \Gamma_{Q+q}^{\alpha\beta}(-Q,Q;-E_{*})+\Gamma_{Q-q}^{\alpha\beta}(-Q,Q;-E_{*}),\nonumber
\end{eqnarray}
then one has $\Gamma_{11}=A^{\alpha\alpha}(q=0)$ 
and $\Gamma_{12}(E)=B^{\alpha\alpha}(q=Q)$, with $\alpha=a$ for
$J_{\perp}>0$ and $\alpha=c$ for
$J_{\perp}<0$. One can rewrite Eqs.\ (\ref{BS-gen}) as
\begin{eqnarray} 
\label{BS-sym}
A_{q}^{\alpha\beta}&=&u_{q}^{\alpha\beta}(0)-\frac{1}{L}\sum_{p}\sum_{\gamma}\frac{u_{q}^{\alpha\gamma}(p)A_{p}^{\gamma\beta}}
{\varepsilon_{Q+p}^{\gamma}+\varepsilon_{Q-p}^{\gamma}+E_{*}},\nonumber\\
B_{q}^{\alpha\beta}&=&v_{q}^{\alpha\beta}(Q)-\frac{1}{L}\sum_{p}\sum_{\gamma}\frac{v_{q}^{\alpha\gamma}(p)B_{p}^{\gamma\beta}}
{2\varepsilon_{p}^{\gamma}+E_{*}},
\end{eqnarray}
where symmetrized kernels $u_{q}^{\alpha\beta}(p)$,
$v_{q}^{\alpha\beta}(p)$ are even functions of $p$:
\begin{eqnarray} 
\label{sym-kernels} 
&&
u_{q}^{\alpha\beta}(p)=\frac{1}{2}\big\{V^{\alpha\beta}_{q-p}(Q+p,Q-p)+V^{\alpha\beta}_{q+p}(Q-p,Q+p)
\big\},\nonumber\\
&&v_{q}^{\alpha\beta}(p)=\frac{1}{2}\big\{V^{\alpha\beta}_{p+q}(-p,p)+V^{\alpha\beta}_{p-q}(-p,p)
\big\}.
\end{eqnarray}

Assume for definiteness that
$J_{\perp}>0$, then of eight equations  (\ref{BS-sym}) we need only a pair of equations for
$A_{q}^{aa}$ and $A_{q}^{ca}$, and another pair of equations for $B_{q}^{aa}$ and $B_{q}^{ca}$.
Solutions to those equations can be
now sought in the form containing only even Fourier harmonics:
\begin{eqnarray} 
\label{ansatz} 
&& A^{aa}_{q}=x_{0}+x_{1}\cos q+ x_{2}\cos 2q,\nonumber\\
&& A^{ca}_{q}=y_{0}+y_{1}\cos q+ y_{2}\cos 2q,\nonumber\\
&& B^{aa}_{q}=\widetilde{x}_{0}+\widetilde{x}_{1}\cos q+ \widetilde{x}_{2}\cos 2q,\\
&& B^{ca}_{q}=\widetilde{y}_{0}+\widetilde{y}_{1}\cos q+ \widetilde{y}_{2}\cos 2q.\nonumber
\end{eqnarray}
This ansatz transforms each of the above pairs of the integral
equations  into a system of 6 linear equations for 6
variables. However, it follows from those equations that
\begin{equation} 
\label{reduce1}
 y_{1}=x_{1},\quad \widetilde{y}_{1}=\widetilde{x}_{1},\quad
y_{2}=x_{2},\quad \widetilde{y}_{2}=\widetilde{x}_{2},
\end{equation}
 so the size of the
corresponding linear problems reduces to $4\times4$. For
$S=\frac{1}{2}$, one has to perform the limit $U\to+\infty$.
After solving the linear
systems, one can read off the effective couplings
\begin{equation} 
\label{result} 
\Gamma_{11}=x_{0}+x_{1}+x_{2},\quad 
\Gamma_{12}=\widetilde{x}_{0}-\frac{\beta}{4}\widetilde{x}_{1}+ \big( \frac{\beta^{2}}{8}-1\big)\widetilde{x}_{2}.
\end{equation}
The procedure for
$J_{\perp}<0$ follows
Eqs.\ (\ref{ansatz})-(\ref{result}), with the obvious interchange
$a\leftrightarrow c$ of
the magnon branch labels.
Below we list the equations for $x_{i}$, $y_{i}$, $\widetilde{x}_{i}$,
$\widetilde{y}_{i}$ in the form that is valid for any sign of
$J_{\perp}$ as well as for any value of $S$, including
$S=\frac{1}{2}$. The resulting systems of equations can be cast into
the following form:
\begin{widetext}
\begin{eqnarray} 
\label{system11}
&& \begin{pmatrix}
\frac{1}{J_{\perp}}+I_{11}^{b} & -\frac{1}{J_{\perp}}-I_{11}^{t} &
I_{12}^{b}-I_{12}^{t} & I_{13}^{b}-I_{13}^{t} \\[2mm]
\frac{1-1/(2S)}{|J_{\perp}|-2J_{Q}}+I_{11}^{b} &
\frac{1-1/(2S)}{|J_{\perp}|-2J_{Q}}+I_{11}^{t} & I_{12}^{b}+I_{12}^{t} &
I_{13}^{b}+I_{13}^{t} \\[2mm]
I_{12}^{b} & I_{12}^{t} & \frac{1}{J_{1}} +I_{22}^{b}+I_{22}^{t} &
I_{23}^{b} + I_{23}^{t} \\[2mm] 
I_{13}^{b} & I_{13}^{t} & I_{23}^{b}+I_{23}^{t} &
\frac{1}{J_{2}}+ I_{33}^{b} + I_{33}^{t} 
\end{pmatrix} 
\begin{pmatrix} x_{0} \\[2mm] y_{0} \\[2mm] x_{1} \\[2mm] x_{2} \end{pmatrix} = \begin{pmatrix}
  1\\[2mm] 1 \\[2mm] 1 \\[2mm] 1\end{pmatrix}, \\[3mm]
\label{system12}
&&
\begin{pmatrix}
\frac{1}{J_{\perp}}+\widetilde{I}_{11}^{b} & -\frac{1}{J_{\perp}}-\widetilde{I}_{11}^{t} &
\widetilde{I}_{12}^{b}-\widetilde{I}_{12}^{t} & \widetilde{I}_{13}^{b}-\widetilde{I}_{13}^{t} \\[2mm]
\frac{1-1/(2S)}{|J_{\perp}|-2J_{Q}}+\widetilde{I}_{11}^{b} &
\frac{1-1/(2S)}{|J_{\perp}|-2J_{Q}}+\widetilde{I}_{11}^{t} & \widetilde{I}_{12}^{b}+\widetilde{I}_{12}^{t} &
\widetilde{I}_{13}^{b}+\widetilde{I}_{13}^{t} \\[2mm]
\widetilde{I}_{12}^{b} & \widetilde{I}_{12}^{t} & \frac{1}{J_{1}} +\widetilde{I}_{22}^{b}+\widetilde{I}_{22}^{t} &
\widetilde{I}_{23}^{b} + \widetilde{I}_{23}^{t} \\[2mm] 
\widetilde{I}_{13}^{b} & \widetilde{I}_{13}^{t} & I_{23}^{b}+I_{23}^{t} &
\frac{1}{J_{2}}+ \widetilde{I}_{33}^{b} + \widetilde{I}_{33}^{t} 
\end{pmatrix} 
\begin{pmatrix} \widetilde{x}_{0} \\[2mm] \widetilde{y}_{0} \\[2mm]
  \widetilde{x}_{1} \\[2mm] \widetilde{x}_{2} \end{pmatrix}
= \begin{pmatrix} 
  2\\[2mm] 2 \\[2mm] 2\cos Q \\[2mm] 2\cos 2Q\end{pmatrix}
\end{eqnarray}
where the matrix coefficients are given by
\begin{eqnarray} 
\label{Iij} 
&& I_{ij}^{b}=\frac{1}{\pi} \int_{0}^{\pi}\frac{f_{i}f_{j}}{2S(J_{Q+p}+J_{Q-p}-2J_{Q})
+E_{*}},
\quad
I_{ij}^{t}= \frac{1}{\pi} \int_{0}^{\pi}\frac{f_{i}f_{j}}{2S(J_{Q+p}+J_{Q-p}+2|J_{\perp}|-2J_{Q})
+E_{*}},\nonumber\\
&& \widetilde{I}_{ij}^{b}=\frac{1}{\pi} \int_{0}^{\pi} \frac{f_{i}f_{j}}{4S(J_{p}-J_{Q})
+E_{*}},
\quad
\widetilde{I}_{ij}^{t}= \frac{1}{\pi} \int_{0}^{\pi}\frac{f_{i}f_{j}}{4S(|J_{\perp}|+J_{p}-J_{Q})
+E_{*}},\nonumber\\
&& f_{1}=1,\quad f_{2}=\cos p,\quad f_{3}=\cos 2p,
\end{eqnarray}
\end{widetext}
and can all be computed analytically in a closed, though somewhat
cumbersome, form. The solutions for $\Gamma_{11}$, $\Gamma_{12}$, which
follow from Eqs.\ (\ref{system11}),
(\ref{system12}), can be obtained with the help of any good computer algebra
system (we used Maple), but are too bulky to be presented here (the result, saved in a
plain ASCII format, is several megabytes large).



\begin{thebibliography}{10.}
\bibitem{noteHs} The expression (\ref{Hs}) is valid if there are no magnon bound
  states, otherwise the
  actual saturation field  will differ from this value by the
  magnon binding energy.
\bibitem{note-superTonks} The effective continuum theory in the strong
coupling regime  $|g_{11}|m\gg1$,  $|g_{12}|m\gg1$
does not fit into the description in terms of the repulsive
Lieb-Liniger model, but rather corresponds to the so-called ``super
Tonks'' Bose gas \cite{Astrakharchik-superTonks} with the Luttinger parameter
$K<1$.  The interested reader is referred to
Ref. \onlinecite{Kolezhuk+12} for further discussion.
\end{thebibliography}

\end{document}